\documentclass[comment, prl,twocolumn,nofootinbib, preprintnumbers, superscriptaddress]{revtex4}

\usepackage{amsmath,amssymb,bm,slashed,braket}
\usepackage{graphicx}
\usepackage{epstopdf}
\usepackage{float}
\usepackage[colorlinks=true,
        linkcolor=blue,
        urlcolor=blue,
        citecolor=green,          
        bookmarks=true,
        bookmarksnumbered=true,
        breaklinks=true,
        pdfpagemode=Fullscreen,
        pdfstartview=FitBH]{hyperref}

\usepackage[normalem]{ulem}

\usepackage{color}

\definecolor{Orange}{cmyk}{0,0.61,0.87,0}
\definecolor{JungleGreen}{cmyk}{0.99,0,0.52,0}
\definecolor{OliveGreen}{cmyk}{0.64,0,0.95,0.40}
\definecolor{Brown}{cmyk}{0,0.81,1,0.60}
\definecolor{RoyalBlue}{cmyk}{0.71,0.53,0,0.12}
\definecolor{Gray}{cmyk}{0,0,0,0.40}
\definecolor{LightPink}{cmyk}{0.0,0.25,0,0}
\definecolor{LLightPink}{cmyk}{0.0,0.10,0,0}
\definecolor{LightBlue}{cmyk}{0.25,0,0,0}
\definecolor{LightGray}{cmyk}{0,0,0,0.2}

\usepackage{xcolor}
\definecolor{gesfpurple}{rgb}{0.47,0.19,0.42}

\definecolor{gesflanse}{rgb}{0.00,0.50,0.50}

\definecolor{gesfblue}{rgb}{0.08,0.42,0.76}

\definecolor{gesfred}{rgb}{1,0,0}

\definecolor{gesfwhite}{rgb}{1,1,1}

\definecolor{gesfblack}{rgb}{0,0,0}

\newcommand{\geqn}[1]{Eq.\,\hypersetup{linkcolor=blue}(\ref{#1})\hypersetup{linkcolor=blue}}
\newcommand{\gfig}[1]{{\hypersetup{linkcolor=violet}Fig.\,\ref{#1}\hypersetup{linkcolor=blue}}}

\allowdisplaybreaks[4]

\newcommand{\titledef}{JUNO and RG running}

\newcommand{\Nuc}[2]{{\ensuremath{\mbox{}^{#1}}\text{#2}}}

\newcommand{\dMa}{\delta_{\rm M1}}

\newcommand{\dMc}{\delta_{\rm M3}}

\newcommand{\dD}{\delta_D}

\newcommand{\TS}{\theta_{s}}
\newcommand{\TR}{\theta_{r}}
\newcommand{\TA}{\theta_{a}}

\newcommand{\cS}{c_{s}}
\newcommand{\sS}{s_{s}}
\newcommand{\tS}{\tan \TS}

\newcommand{\cR}{c_{r}}

\newcommand{\tR}{\tan \TR}

\newcommand{\DdD}{\Delta\delta_D}

\newcommand{\DTS}{\Delta\theta_{s}}
\newcommand{\DTR}{\Delta\theta_{r}}

\newcommand{\DdMa}{\Delta\delta_{\rm M1}}

\newcommand{\DdMc}{\Delta\delta_{\rm M3}}

\newcommand{\BTS}{\beta_{s}}
\newcommand{\BTR}{\beta_{r}}

\newcommand{\BdD}{\beta_{D}}
\newcommand{\BdMa}{\beta_{\mathrm{M}1}}

\newcommand{\BdMc}{\beta_{\mathrm{M}3}}

\newcommand{\Dms}{\Delta m^2_{21}}

\newcommand{\DP}{\Delta\Phi}
\newcommand{\DPee}{\Delta\Phi_{ee}}
\newcommand{\DT}{\Delta\Theta}

\newcommand{\suppsection}[1]{
  \vspace{0.4cm}              
  \noindent{\bfseries #1}
  \vspace{0.25cm}             
  \par\nopagebreak            
}

\hypersetup{ pdfauthor = {Shao-Feng Ge},
pdftitle = {\titledef},
pdfsubject = {},
pdfkeywords = {},
pdfcreator = {LaTeX with hyperref package},
pdfproducer = {dvips + ps2pdf} }

\usepackage{stackengine}

\definecolor{Orange}{cmyk}{0,0.61,0.87,0}
\definecolor{JungleGreen}{cmyk}{0.99,0,0.52,0}
\definecolor{OliveGreen}{cmyk}{0.64,0,0.95,0.40}
\definecolor{Brown}{cmyk}{0,0.81,1,0.60}
\definecolor{RoyalBlue}{cmyk}{0.71,0.53,0,0.12}
\definecolor{Gray}{cmyk}{0,0,0,0.40}
\definecolor{LightPink}{cmyk}{0.0,0.25,0,0}
\definecolor{LLightPink}{cmyk}{0.0,0.10,0,0}
\definecolor{LightBlue}{cmyk}{0.25,0,0,0}
\definecolor{LightGray}{cmyk}{0,0,0,0.2}

\definecolor{byzantine}{rgb}{0.74, 0.2, 0.64}
{}

\usepackage{float}

\graphicspath{{figs/}}

\usepackage{subfiles}

\begin{document}

\title{Disentangle RG Running Parameters with Medium-Baseline Reactor Experiments}

\author{Shao-Feng Ge}
\email{gesf@sjtu.edu.cn}
\affiliation{State Key Laboratory of Dark Matter Physics, Tsung-Dao Lee Institute \& School of Physics and Astronomy, Shanghai Jiao Tong University, China}
\affiliation{Key Laboratory for Particle Astrophysics and Cosmology (MOE) 
\& Shanghai Key Laboratory for Particle Physics and Cosmology, Shanghai Jiao Tong University, Shanghai 200240, China}

\author{João Paulo Pinheiro}
\email{joaopaulo.pinheiro@sjtu.edu.cn}
\affiliation{State Key Laboratory of Dark Matter Physics, Tsung-Dao Lee Institute \& School of Physics and Astronomy, Shanghai Jiao Tong University, China}
\affiliation{Key Laboratory for Particle Astrophysics and Cosmology (MOE) 
\& Shanghai Key Laboratory for Particle Physics and Cosmology, Shanghai Jiao Tong University, Shanghai 200240, China}

\author{Shaoyang Qin}
\email{shaoyang.qin@campus.lmu.de}
\affiliation{State Key Laboratory of Dark Matter Physics, Tsung-Dao Lee Institute \& School of Physics and Astronomy, Shanghai Jiao Tong University, China}
\affiliation{Key Laboratory for Particle Astrophysics and Cosmology (MOE) 
\& Shanghai Key Laboratory for Particle Physics and Cosmology, Shanghai Jiao Tong University, Shanghai 200240, China}
\affiliation{Arnold-Sommerfeld-Center for Theoretical Physics, Ludwig-Maximilians-Universität München, Theresienstraße 37, 80333 München, Germany}

\begin{abstract}

We study how the renormalization group running beta functions
of mixing angles and leptonic CP phases affect
the slow and fast oscillation modes at JUNO. While the slow
mode is modulated by the solar parameters, its amplitude
can also be affected by the solar angle beta function $\beta_s$
and its phase by the Majorana CP phase counterpart $\beta_{\rm M1}$.
On the other hand, the fast mode also receives corrections
from the beta functions of the Dirac CP phase $\delta_D$
and the Majorana CP phase $\delta_{\rm M3}$. Since the
fast mode is essentially the one measuring the neutrino
mass ordering, the RG running effect can then interfere
to deteriorate the sensitivity. Fortunately, the JUNO-TAO
near detector can provide supplementary measurements of
the RG running parameters to restore the mass ordering
sensitivity. Since the rephasing phases $\delta_{\rm M1}$ and $\delta_{\rm M3}$ can be
shifted by a vector-like rephasing for Dirac neutrinos, their running is
physical only in the Majorana case. In consolidated,
the mild preference for a nonzero $\beta_{\rm M1}$ by
the current JUNO data thus hints the
Majorana nature of neutrinos.

\end{abstract}
\maketitle

\textbf{Introduction} --
The three-flavor framework of neutrino oscillations is 
firmly established \cite{ParticleDataGroup:2024cfk}
with the two mass-squared splittings and three
mixing angles being determined to the few-percent level
\cite{Esteban:2024eli,Capozzi:2025wyn,deSalas:2020pgw}. Nevertheless,
two oscillation observables remain unresolved: The neutrino 
mass ordering (MO) and the Dirac CP phase $\dD$.
Beyond the oscillation sector lies
an even more fundamental question: whether neutrinos are Dirac or Majorana
particles. For Majorana neutrinos, there would be
two additional CP phases which in the standard 
picture are altogether inaccessible to oscillation experiments \cite{Schechter:1981bd, Dolinski:2019nrj,Agostini:2022zub, Cirigliano:2022oqy}.

The MO is being scrutinized along complementary directions.
First, the matter resonance effects
\cite{Akhmedov:2012ah, Ribordy:2013xea, Ge:2013zua, Ge:2013ffa, IceCubeCollaboration:2023wtb} in the atmospheric
and long-baseline accelerator experiments
are sensitive to the ordering-dependent modification on the oscillation 
probabilities \cite{Petcov:2001sy,Choubey:2003qx,Nunokawa:2005nx,Minakata:2006gq, Minakata:2007tn,Blennow:2013oma,Parke:2024xre}. In addition, the vacuum-dominated reactor channel 
determines MO through the oscillation phases \cite{Petcov:2001sy,Choubey:2003qx,Nunokawa:2005nx,Minakata:2006gq,Minakata:2007tn,Zhan:2008id,Zhan:2009rs}.
With a $20$\,kton liquid-scintillator detector at
a medium baseline of $\sim 53$\,km and
unprecedented energy resolution of $\sim 3\%/\sqrt{E{\rm /MeV}}$,
JUNO can resolve the fast atmospheric wiggles
riding on the slow solar modulation \cite{Qian:2012xh,Ge:2012wj,JUNO:2015zny}.
Since data collection in the summer of 2025,
JUNO has already published the first data for the slow
mode \cite{JUNO:2025gmd} and the second data for the fast mode
\cite{JUNO:2026talk}.
However, the leptonic CP phases do not appear in the
electron-antineutrino survival probability $P_{\bar e \bar e}$
that is measured by the reactor experiments. 

The whole neutrino oscillation process contains not just
the oscillation part but also the production and detection processes.
With two interaction vertices, the oscillation amplitude
contains two mixing matrices, each with a set of mixing
parameters. The cancellation and absence of both Dirac
and Majorana CP phases in $P_{\bar e \bar e}$
are consequences of the fact
that the two mixing matrices are actually the same.
If the two mixing matrices differ from each other, for
example, due to the renormalization group (RG) running,
the leptonic Dirac CP phase can have phenomenological
consequences \cite{Babu:2021cxe,Ge:2023azz} even in the reactor
neutrino oscillation experiments \cite{Ge:2024ibn}.

We would like to emphasize that the appearance of
the leptonic CP phase RG running parameters applies
not just to the Dirac one but also to those diagonal
rephasing phases associated with neutrino fields.
Although the diagonal rephasing phases are hidden even deeper
with complete absence from all the oscillation probabilities,
the RG running effect would also bring their beta functions
into play. 
However, the running of these diagonal rephasing phases
is physical only for Majorana neutrinos. For Dirac neutrinos,
these diagonal phases can be fully reabsorbed by a
vector-like rephasing of the left- and
right-handed fields, leaving no imprint on $P_{\bar e \bar e}$.
A nonzero $\BdMa$ or $\BdMc$  is therefore a purely Majorana signal,
providing an unexpected way to distinguish the two cases at 
reactor neutrino oscillation experiments.
We systematically investigate the RG
running effects for the JUNO experiment. Especially,
the leptonic CP phases can interfere with the
MO measurement and the JUNO-TAO near detector
plays a key role of restoring the MO sensitivity.

\vspace{2mm}
\textbf{Neutrino Oscillation with RG Running}
--
In the presence of RG running, the two mixing matrices
$U(Q^2_p)$ and $U(Q^2_d)$ for the production and
detection processes, respectively,
could differ with mismatched momentum transfers
$Q^2_p$ and $Q^2_d$. The oscillation probability
\cite{Babu:2021cxe,Ge:2023azz},
\begin{equation}
  P_{\alpha\beta}
=
\left|
  \sum_{i=1}^3 U_{\beta i}(Q^2_d)
  \exp \left( -i \dfrac{m_i^2 L}{2 E_\nu} \right) 
  U^*_{\alpha i}(Q^2_p)
\right|^2,
\label{eq:RG_oscillation_probability}
\end{equation}
where $L$ is the baseline and $E_\nu$ is
the neutrino energy, can then develop effective
non-unitarity \cite{Ge:2024ibn}. In the
zero-distance limit ($L \rightarrow 0$),
the effective non-unitarity
$U(Q^2_d) U^\dagger(Q^2_p) \neq \mathbb I$
leads to nonzero flavor transition
probabilities $P_{\alpha \beta}(L \rightarrow 0) \neq 1$
for $\alpha \neq \beta$ \cite{Babu:2021cxe,Ge:2023azz}.

The RG running of neutrino mixing parameters
can be parametrized with beta functions $\beta_{X_I}$,
\begin{equation}
  \Delta X_I
=
  \beta_{X_I} \ln \left(\left|\dfrac{Q_d^2}{Q_p^2}\right|\right),
\quad
  \beta_{X_I} \equiv \dfrac{\partial X_I}{\partial\ln \mu^2},
\label{eq:RG_log}
\end{equation}
where $X_I \equiv \{\TS,\TA,\TR, \dD,\dMa,\dMc \}$ and
$\mu^2$ is the renormalization scale. 
For clarity, we have denoted the three mixing angles as
$\TS \equiv \theta_{12}$ for the solar ($s$) angle, 
$\TR \equiv \theta_{13}$ for the reactor ($r$) angle, and
$\TA \equiv \theta_{23}$ for the atmospheric ($a$) angle
according to the corresponding major experiments
that measured their values.

Expanding \geqn{eq:RG_oscillation_probability} with 
this factorization  yields, the $\bar\nu_e$ survival
probability in vacuum reads, 
\begin{align}
  P_{\bar{e}\bar{e}}
\approx
  1
& - 4 \cR^4 c^2_s s^2_s (1+ \DT_{21}) \sin^2 \left(\Phi_{21}+ \DP_{21} \right)
\notag \\
& - 4 c^2_r s^2_r \cS^2(1 + \DT_{31}) \sin^2 \left(\Phi_{31}+ \DP_{31} \right)
\notag \\
& - 4 c^2_r s^2_r \sS^2 (1 + \DT_{32}) \sin^2 \left(\Phi_{32}+ \DP_{32} \right),
\label{eq:Pee_RG_full}
\end{align}
where $\Phi_{ij}\equiv \Delta m_{ij}^2 L/4E_\nu$ 
is the kinematic oscillation phases corresponding to
the mass squared difference $\Delta m^2_{ij} \equiv m^2_i - m^2_j$. 
The RG running effects $\Delta \theta_s$ and $\Delta \theta_r$
of mixing angles have been summarized
in the amplitude corrections,
\begin{subequations}
\begin{align}
  \DT_{21}
& \equiv
  (\cot\TS-\tS)\DTS 
- 2 \tR \DTR,
\\
  \DT_{31}
& \equiv
- \tS \DTS 
+ (\cot \TR - \tR)\DTR,
\\
 \DT_{32}
& \equiv
  \cot \TS \DTS
+ (\cot \TR - \tR) \DTR,
\end{align}
\label{eq:def_omega}
\end{subequations}
from one of the two mixing matrices in 
\geqn{eq:RG_oscillation_probability}.
However, the solar and reactor mixing angles typically
receive hierarchical RG running effects as inherited
from the hierarchy in the two mass squared differences.
Being dimensionless, the mixing angle $\theta_{ij}$ and
its beta function $\beta_{ij}$ is often inversely proportional to
the corresponding mass squared difference $\Delta m^2_{ij}$.
With roughly the same size of radiative corrections to the
mass matrix, the mixing angle beta functions roughly scales
as, $\BTR / \BTS \propto \Delta m^2_{21} / \Delta m^2_{31}
\approx 3\%$ \cite{Ohlsson:2013xva}. Then the $\DTR$ terms above
can be safely omitted.

The PMNS mixing matrix $U \equiv \mathcal P V \mathcal Q$
contains two diagonal rephasing matrices $\mathcal P$ and
$\mathcal Q$. While $\mathcal P$ is purely unphysical,
$\mathcal Q \equiv \mbox{diag} \{e^{i \dMa / 2}, 1, e^{i \dMc / 2} \}$
contains two Majorana CP phases $\dMa$ and $\dMc$. 
Even for Dirac neutrinos, $\mathcal Q$ still appears in the
parameterization, although the two phases are then unphysical.
The left-handed mixing matrix $U_L$ of neutrinos entering
\geqn{eq:RG_oscillation_probability} and its right-handed
counterpart $U_R$ diagonalize $M_\nu M_\nu^\dagger$ and
$M_\nu^\dagger M_\nu$, respectively, with the neutrino mass matrix being
$M_\nu = U_L\,\mathrm{diag}(m_i)\,U_R^\dagger$. For each mixing
matrix, it is possible to parametrize in the similar way as
the PMNS matrix,
$U_L \equiv \mathcal P_L V_L \mathcal Q_L$ and 
$U_R \equiv \mathcal P_R V_R \mathcal Q_R$. Then the diagonal
mass matrix $\mathrm{diag}(m_i)$ is sandwiched by
the two diagonal rephasing matrices $\mathcal Q_L$
and $\mathcal Q^\dagger_R$. Since all three of $\mathcal Q_L$,
$\mathrm{diag}(m_i)$, and $\mathcal Q_R^\dagger$ are diagonal,
they commute with each other and the two rephasing
matrices $\mathcal Q_L$ and $\mathcal Q^\dagger_R$ merge.
There is no practial way to distinguish $\mathcal Q_L$
and $\mathcal Q^\dagger_R$. Only the difference between the
CP phases ($\delta_{\rm M1}$ and $\delta_{\rm M3}$)
of $\mathcal Q_L$ and $\mathcal Q_R$ can finally
appear in the neutrino mass matrix \cite{Huang:2025qew}.
It is then perfectly fine to shift $\delta_{\rm M1}$
and $\delta_{\rm M3}$ away from $\mathcal Q_L$ to $\mathcal Q_R$
leaving $\mathcal Q_L = \mathbb I_{3 \times 3}$. 
For Majorana neutrinos, $M_\nu$ is symmetric and 
$U_R=U_L^*$ and hence those phases of $\mathcal Q_L$
(or simply $\mathcal Q$ since $\mathcal Q_L$ and
$\mathcal Q_R$ are actually the same now)
cannot be removed.

Then one may see from \geqn{eq:RG_oscillation_probability}
that the phases of $\mathcal P$ always disappear when
taking absolution values since they contribute as
overall multiplicative factors to the quantum mechanical
oscillation amplitude. However, any mismatch in the
$\mathcal Q$ phases can survive and hence contribute
to those phasing shifts $\Delta \Phi_{ij}$ in 
\geqn{eq:Pee_RG_full},
\begin{subequations}
\begin{align}
  \DP_{21} & \equiv - \tfrac 1 4 \DdMa,
\\ 
  \DP_{32} & \equiv - \tfrac 1 2 \DdD + \tfrac 1 4 \DdMc,
\\
  \DP_{31} & \equiv - \tfrac 1 2 \DdD - \tfrac 1 4 \DdMa + \tfrac 1 4 \DdMc.
\end{align}
\label{eq:def_upsilon}
\end{subequations}
Note that the Dirac CP phase $\delta_D$ can also shift
the oscillation phases since it appears in the $U_{e 3}$
element as an overall phase.

While the first term of \protect\geqn{eq:Pee_RG_full}
is the \textit{slow} oscillation governed by the solar
mass splitting $\Dms$, the second and third are the \textit{fast}
oscillation controlled  by the atmospheric splittings
$\Delta m^2_{31}$ and $\Delta m^2_{32}$, respectively.
It is then possible to use the slow and fast oscillation
modes at reactor experiments to probe the corresponding
RG running parameters.

\vspace{2mm}
{\bf Slow Oscillation Mode with $\dMa$ and $\theta_s$ RG Running}
--  
At medium-baseline experiments, the unsuppressed
slow term determines the
overall modulation across the few-MeV range, with
its amplitude and frequency encoding the solar
mixing angle $\TS$ and mass squared difference
$\Dms$, respectively. The RG running introduces two distinct
distortions by the beta function $\BTS$
that alters the amplitude and
$\BdMa$ that produces an energy-dependent
phase drift.

Since the running enters through the detection momentum
transfer, the survival probability is the
cross-section-weighted average over 
$Q_d^2$ \cite{Ge:2024ibn}.

\begin{figure}[t]
\centering
\includegraphics[width=1\linewidth]{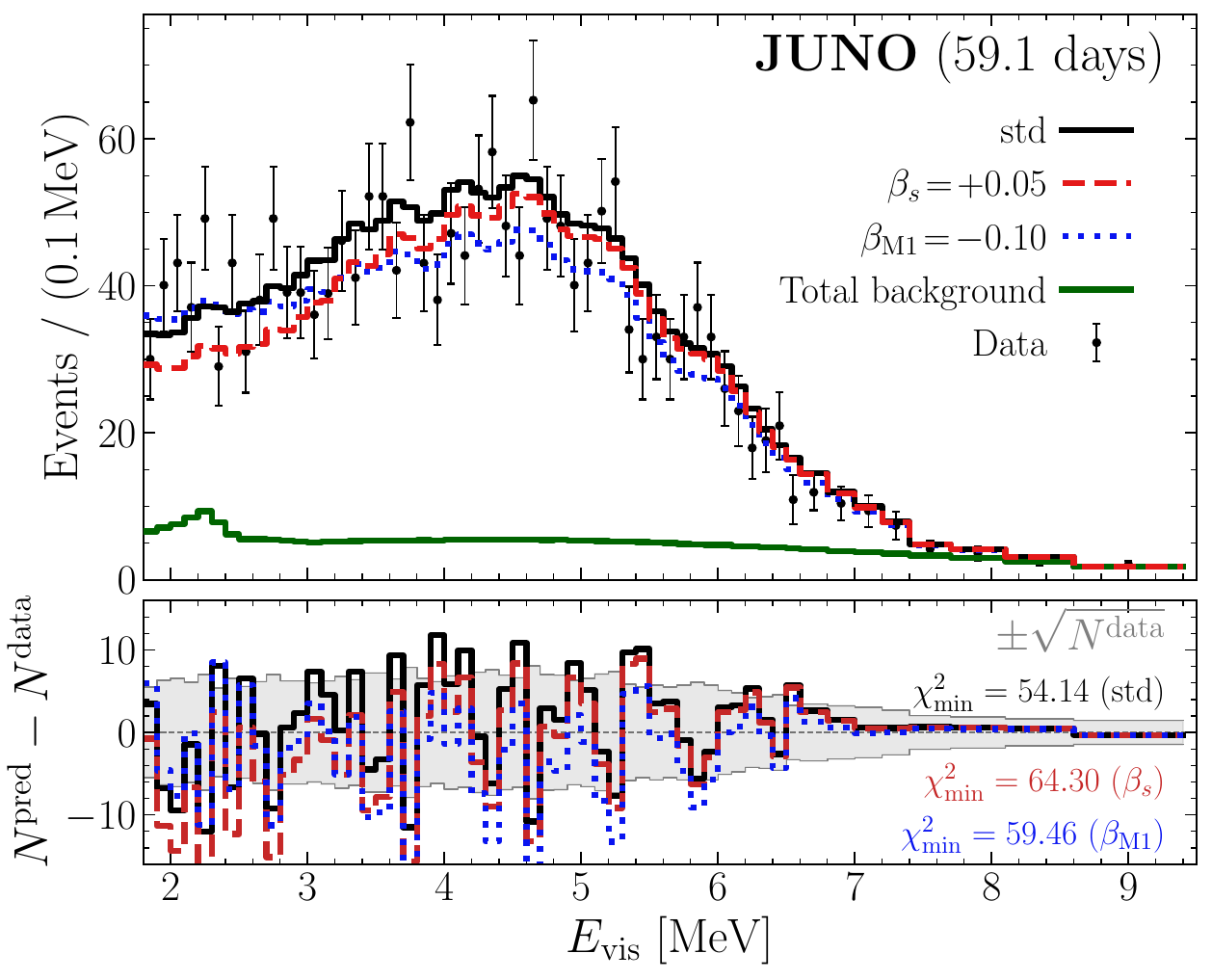}
\caption{%
(Upper) The IBD events from the first data release of JUNO
as a function of visible energy
$E_{\rm vis}$ together with the total background (green).
For comparison, the best-fit results for the standard
three-neutrino oscillation (black), the RG running case
with $\beta_s = 0.05$ (red dashed) and $\beta_{\rm M1} = -0.1$
(blue dotted) are shown together.
(Lower) The residual $N^{\rm pred}_i-N^{\rm data}_i$
normalized to $\pm\sqrt{N^{\rm data}_i}$.}
\label{fig:events_rg_solar}
\end{figure}

\gfig{fig:events_rg_solar} shows the IBD event spectrum
for the first data release of JUNO
\cite{JUNO:2025gmd}. 
For comparison, we have also shown the best-fit
curves of the standard three-neutrino mixing (black solid)
and the case with RG runnings.
A non-zero $\BTS$ rescales the amplitude
of the event spectrum, much like a shift in the solar angle
$\TS$.
Although the event rate is not that large in the high-energy
end, the corresponding relative error there is actually smaller
as indicated by the lower panel.
Consequently, the best-fit curves of the three different cases
are constrained to be almost the same for $E_{\rm vis} \gtrsim 6$\,MeV.
With this anchor, the largest deviations seem to appear in
the low-energy region for the $\beta_s = 0.05$ case (red
dashed) and the intermediate region for $\beta_{\rm M1} = -0.1$
(blue dotted).
Such distinct energy fingerprints of $\BTS$ and $\BdMa$
follow directly from the kinematic structure of the slow term.
Entering through the amplitude, the $\BTS$ distortion has
the largest effect when the solar dip is deepest at $\Phi_{21}=\pi/2$,
i.e., $E_\nu\approx 3.2$\,MeV ($E_{\rm vis}\approx 2.4$\,MeV).
For comparison, $\beta_{\rm M1}$ enters through the phase
correction with its coefficient scaling as $\sin 2\Phi_{21}$
and hence its effect maximizes at
$\Phi_{21}=\pi/4$ ($E_{\rm vis }\approx 5$ MeV).
However, it does not mean that the energy-dependent RG effect
has smaller energy dependence at the high-energy region.

\begin{figure}[t]
\centering
\includegraphics[width=1\linewidth]{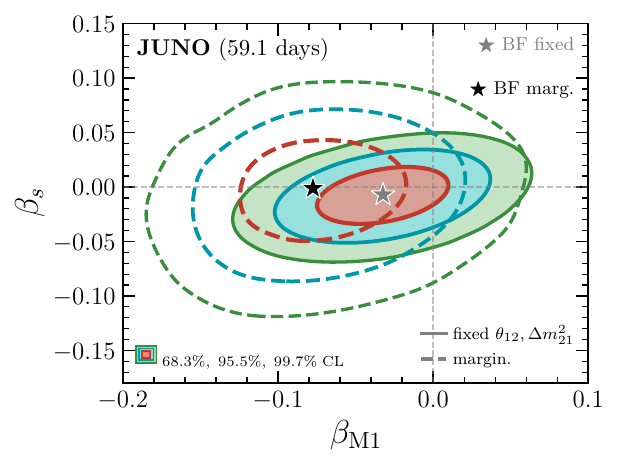}
\caption{The JUNO sensitivity in the $(\BTS,\BdMa)$ plane at
$1,2,3\sigma$ C.L. with the standard solar  parameters fixed
at the NuFIT~6.1 best fit (solid) or marginalized over the
standard solar parameters with NuFIT~6.0 before JUNO (dashed).}
\label{fig:2d_bt12_bdm1}
\end{figure}

\gfig{fig:2d_bt12_bdm1} shows the resulting exclusion
contours in the $(\BTS,\BdMa)$ plane.
The slightly tilted $1\sigma$ contour shows that the 
two parameters are weakly correlated
and the slight asymmetry along the
$\BTS$ axis reflects the non-linear dependence of the event
rate on $\sin^2 \theta_s$.
The first JUNO data release mildly favors a nonzero
Majorana-phase running at the $1\sigma$ level. Along
the $\BTS=0$ slice, the minimum lies at $\BdMa=-0.078$ with
$\chi^2_{\rm std}-\chi^2_{\rm RG}=3.8$ relative to the 
standard  ($\beta=0$) solution. 
Fixing the solar parameters to the NuFIT 6.1
best-fit, the $90\%$ C.L. intervals  (1~d.o.f.) are
$\BTS\in[-0.036,+0.020]$ and $\BdMa\in[-0.078,+0.013]$.
Already with the second
data release \cite{JUNO:2026talk} ($207.2$ days), the 
corresponding limits improve to $\BTS \in [-0.016,+0.025] $ and $\BdMa \in [-0.073,-0.010]$ at $90\%$\,C.L.. 
Since $\BdMa$ can run only for Majorana neutrinos, 
this consistent pull towards a nonzero $\BdMa$ across both
data releases translates into a mild JUNO preference for the Majorana
nature with RG running and its significance
is expected to further enhance with the
projected exposure.

Looking ahead to the projected exposure, we repeat the analysis
on an Asimov data set for the full JUNO \cite{Forero:2021lax}
generated under the normal ordering (NO) with the 
NuFIT 6.1 best-fit parameters \cite{nufit61} for the standard oscillation
scenario. After $2400$ effective
detector days, the projected $90\%$\,C.L.\ bounds improve to
$|\BTS|<0.0066$ and $|\BdMa|<0.012$. JUNO can
constrain the RG running of both solar parameters at the 
percent-to-permille level.

\vspace{2mm}
{\bf Fast Oscillation Mode with RG Running of CP Phases}  
---
The essential measurement of the neutrino mass ordering at JUNO
can be summarized into the fast mode
$P^{\rm fast}_{\bar{e}\bar{e}} \propto \cos\!\big(2\Phi_{ee}\pm\Phi_\odot\big)$
with $\Phi_{ee}\equiv \Delta m^2_{ee} L / 4 E_{\nu}$
\cite{Minakata:2007tn,JUNO:2015zny}.
While the effective
solar phase $\Phi_\odot\equiv \arctan(\cos2\TS \tan\Delta_{21})-\Delta_{21}\cos2\TS$
is always positive, the oscillation phase
$\Phi_{ee}$ flips sign between the two mass orderings.
The effective argument then
\textit{advances} for NO and \textit{retards} for IO.
Note that this analytical way is consistent with and
can directly connect with the Fourier method
\cite{Zhan:2008id,Zhan:2009rs}.

The RG running enters as shifting the apparent atmospheric splitting
$\Delta m^2_{ee}\equiv \cS^2 \Delta m^2_{31} +\sS^2 \Delta m^2_{32}$
by $-\,(4E_\nu/L)\,\DPee$ with,
\begin{equation}
  \DPee
\equiv
  \tfrac 1 2 \DdD
+ \tfrac 1 4 \cS^2 \DdMa
- \tfrac 1 4 \DdMc.
\label{eq:DeltaTheta_def}
\end{equation}
For Dirac neutrinos,  both $\DdMa$ and $\DdMc$
can be absorbed such that the fast mode correction
$\DPee = \tfrac12\DdD$ is driven by the Dirac-phase
running $\BdD$ alone. The correction $\Delta \Phi_{ee}$
could be nonzero for both Dirac and Majorana neutrinos. 
Unlike the slow mode, however, the fast mode
only probes this single combination and cannot by itself separate
$\BdD$ from $\BdMc$. The Dirac/Majorana discrimination therefore
rests entirely on the slow-mode $\BdMa$.
In addition, the slow-mode correction can also affect the
effective oscillation phase,
\begin{subequations}
\begin{align}
\text{NO:}\quad & \propto\cos\big(2|\Phi_{ee}|-2\DPee+\Phi_\odot  - \Delta\Phi_\odot\big)\,, \\
\text{IO:}\quad & \propto\cos\big(2|\Phi_{ee}|+2\DPee-\Phi_\odot  + \Delta\Phi_\odot\big)\,,
\end{align}
\label{eq:phase_advance_RG}
\end{subequations}
with
\begin{align}
\Delta\Phi_\odot \equiv
& \ \frac {\sin 2\TS \sin\Phi_{21}}
          {\cos^2\Phi_{21} + \cos^2 2\TS \sin^2\Phi_{21}}
\Big[\cos\Phi_{21}\,\DTS
\nonumber \\
& \hspace{14mm}
+ \tfrac 1 4 \cos 2\TS \sin 2\TS \sin\Phi_{21} \DdMa \Big] \,.
\end{align}
Although $\Phi_{ee}$ flips sign, the sign in the cosine
function can be shifted to the other terms $\Delta \Phi_{ee}$,
$\Phi_\odot$, and $\Delta\Phi_\odot$ that does not flip sign.
The RG running affects the fast oscillation wiggles
through $\DPee$ by shifting the apparent $\Delta m^2_{ee}$
or $\Phi_{ee}$. On the other hand, the slow-mode term
$\Delta\Phi_\odot$ shifts the solar phase $\Phi_\odot$.

Note that $\DTS$ and $\DdMa$ in $\Delta\Phi_\odot$
and $\Delta \Phi_{ee}$
also appear in the slow mode and hence can already be pinned
at the $\mathcal{O}(10^{-2}\!-\!10^{-3})$ level with the full
JUNO experiment. The slow and fast modes are
therefore \emph{complementary}.

\begin{figure}[t]
\centering
\includegraphics[width=1\linewidth]{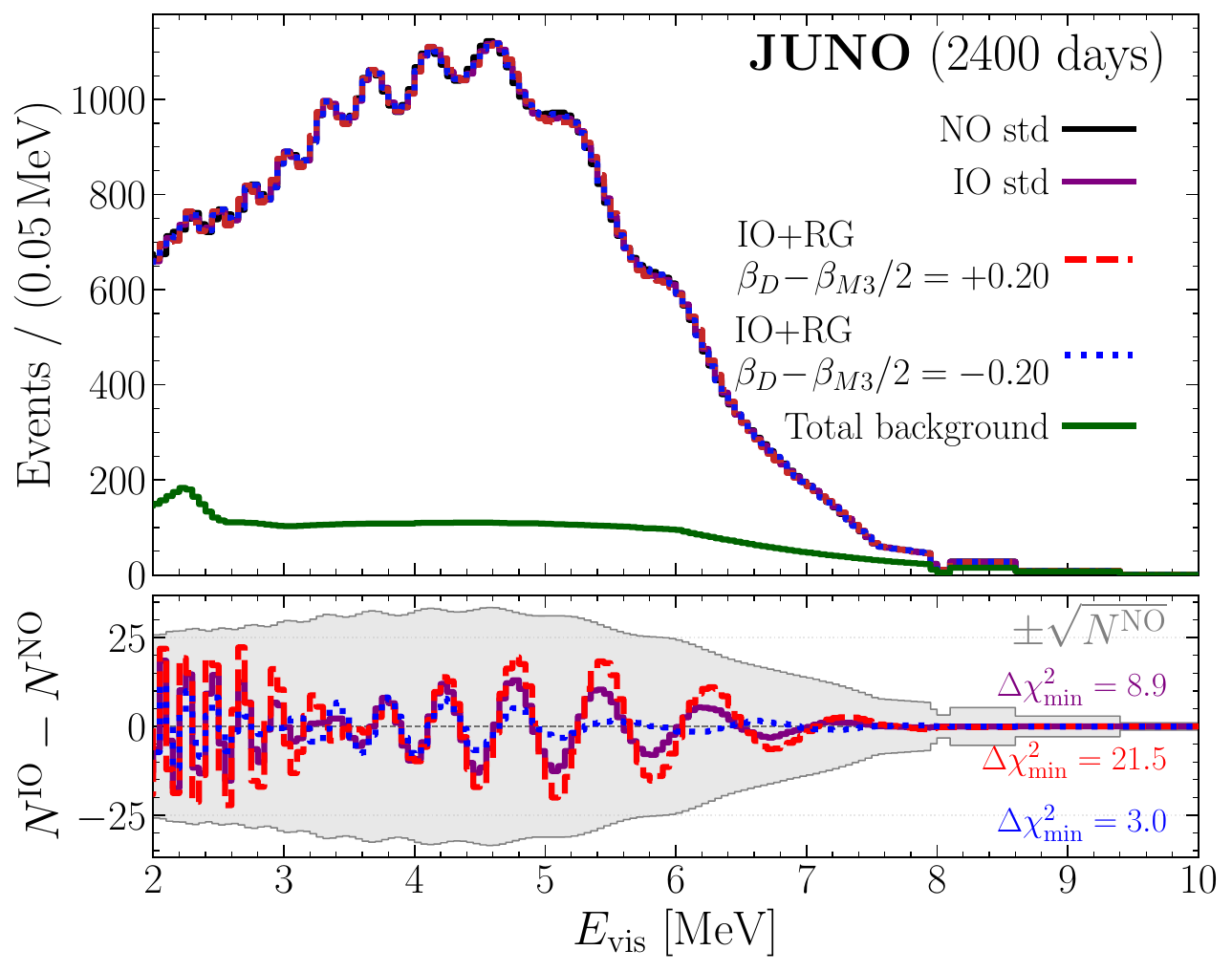}
\caption{%
(Upper)
The expected IBD event spectrum at JUNO versus visible energy, for 2400 days of operation with the standard NO (black solid)
or IO (purple solid). Curves with RG running
$\BdD - \BdMc / 2 = 0.2$ (red dashed)
and $\BdD - \BdMc/2 = -0.2$ (blue dotted)
are also shown for comparison while the green curve stands
for the total background.
(Lower) The residual deviations $N^{\rm IO}-N^{\rm NO}$ and a grey
band encompassing the statistical fluctuations $\pm\sqrt{N^{\rm NO}}$.}
\label{fig:events_mo_comparison}
\end{figure}

\gfig{fig:events_mo_comparison} shows the standard NO
and IO event spectra as well as the IO spectra with
the RG running effect $\BdD - \BdMc/2 = \pm0.2$.
Taking NO with
$\Delta m_{ee}^2 = 2.52\times10^{-3}$\,eV$^2$ as the standard case,
the full JUNO with 2400 days can reach $\Delta\chi^2_{\rm min}({\rm IO})=8.9$,
with the IO minimum located at $\Delta m_{ee}^2\simeq2.54\times10^{-3}$\,eV$^2$.
A negative shift $\BdD - \BdMc/2 = -0.2$ partially erases the
discrimination: it mimics the IO phase retardation,
brings the RG-corrected IO spectrum close to the
standard NO pattern, and hence reduces the sensitivity by a
factor of three to $\Delta\chi^2_{\rm min}({\rm IO})=3.0$. Conversely,
$\BdD - \BdMc/2 = +0.2$ advances the phase further such that the orderings
become easier to separate with $\Delta\chi^2_{\rm min}({\rm IO})=21.5$.
Including the RG running effect, the mass ordering discrimination at
full JUNO weakens from $3.0\sigma$ to $1.6\sigma$ with
$\Delta\chi^2_{\rm min}=2.5$.

\vspace{2mm}
{\bf Guarantee MO Sensitivity with JUNO-TAO}
--
The degeneracy between the mass ordering (or equivalently
$\Phi _{ee}$) and the RG running effect arises because
JUNO measures a single finite-baseline spectrum. While $\Phi_{ee}$
depends on the oscillation baseline and the neutrino energy,
the RG running parameters $\beta_D$ and $\beta_{\rm M3}$
are baseline and energy independent. An extreme supplement
is measuring the neutrino flavor transition \cite{Ge:2024ibn}
at near detectors such as JUNO-TAO \cite{JUNO:2020ijm,JUNO:2024jaw}.

In the RG framework, the near-zero distance
limit  of \geqn{eq:Pee_RG_full} retains a residual $\mathcal{O}(\Delta^2)$ term
proportional to the \textit{same} phase combination
\cite{Babu:2021cxe,Ge:2023azz},
\begin{equation}
  P_{\bar{e}\bar{e}}(L\to 0)
\approx
  1
- \tfrac{1}{4}\, \sin^2(2\TR) (\DdD-\tfrac{1}{2} \DdMc)^2,
\label{eq:Pee_zero}
\end{equation}
assuming vanishing $\DTS$ and $\DdMa$ that have already been
well constrained by the slow oscillation mode, 
and an also vanishing
$\DTR$ as naturally suppressed in model buildings.
With vanishing baseline, the oscillation phase $\Phi_{ee}$
disappears and only the combination $\beta_D - \beta_{\rm M3} / 2$
that also appears in $\Delta \Phi_{ee}$ retains,
which for Dirac neutrinos reduces to $\BdD$ since
$\BdMc$ can be rephased away.

\begin{figure}[t]
\centering
\includegraphics[width=1\linewidth]{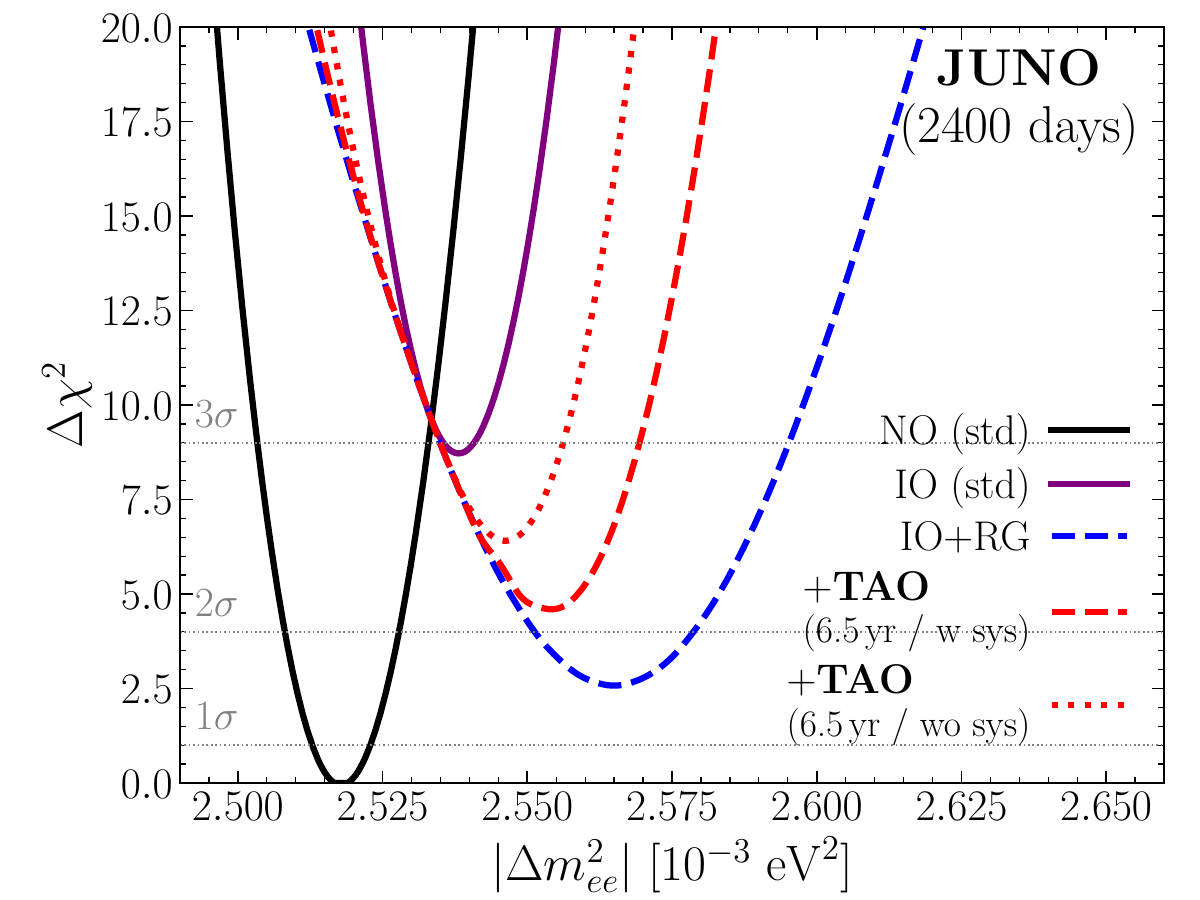}
\caption{The mass ordering sensitivity versus the fitting parameter
$|\Delta m^2_{ee}|$ for the full JUNO far detector alone
(2400 days $\sim 6.5$ years) and for the JUNO+TAO combination.
For comparison, we have shown the fitting with the standard NO
(black solid) or IO (purple solid) as well as the IO fitting supplemented
by RG running with just the full JUNO far detector (blue dashed)
or with both far and near detectors (red).}
\label{fig:ord_rg_chisq}
\end{figure}

\gfig{fig:ord_rg_chisq} shows the mass ordering sensitivity
as a function of the fitting parameter $|\Delta m^2_{ee}|$.
With only the full JUNO far detector, the RG running effect
can reduce the MO sensitivity from $3\sigma$ to
$\Delta\chi^2_{\rm IO}=2.5$ ($1.6\,\sigma$). Adding the
$6.5$-year TAO constraints \cite{Ge:2024ibn} recovers
most of this loss.
In the idealized limit of a perfectly known
normalization (red dotted), the degeneracy is fully broken and the sensitivity
recovers to $\Delta\chi^2_{\rm IO}=6.8$ ($\simeq2.6\,\sigma$).
On the other hand, for a realistic
normalization (red dashed) it is partially broken, still recovering to
$\Delta\chi^2_{\rm IO}=4.9$ ($\simeq2.2\,\sigma$).

\vspace{2mm}
\textbf{Conclusion}
--
The RG running between the neutrino production and
detection processes 
allows reactor experiments
to probe the leptonic CP properties.
We have shown that the RG running
of CP phases $\dD$, $\dMa$, and $\dMc$ has physical consequences
on the neutrino oscillation probabilities,
with the later two being physical only for Majorana neutrinos.
While the beta functions $\beta_s$ and $\delta_{\rm M1}$
of the solar mixing angle and the first Majorana CP phase
affect both the slow and fast oscillation modes,
the beta functions $\beta_D$ and $\beta_{\rm M3}$
of the Dirac and the third Majorana CP phases $\delta_{\rm M3}$
can also modify the fast mode. The two oscillation
modes then provide complementary measurements.

The slow mode constrains the solar running parameters $\BTS$ and
$\BdMa$. 
The first $59.1$-day dataset already mildly favors a nonzero
$\BdMa = -0.078$ at more than $90\%$~C.L\@.\
The second $207.2$-day dataset likewise prefers a nonzero $\BdMa$,
confining it to $[-0.073,\,-0.010]$ at $90\%$~C.L..
Since $\BdMa$ runs only for Majorana neutrinos, 
this persistent preference already points, albeit mildly,
to the Majorana nature.
The nominal $2400$-day exposure is projected to reach bounds at the
$\mathcal{O}(10^{-2}\!-\!10^{-3})$ level for both $\BTS$ and $\BdMa$. 
By contrast, the fast mode is governed by the combination
$\BdD - \BdMc/2$, which shifts the apparent $|\Delta m^2_{ee}|$ and
significantly degrades the JUNO mass-ordering sensitivity from
$3.0\,\sigma$ to $1.6\,\sigma$. Fortunately, in the zero-distance limit
the JUNO-TAO near detector can uniquely probe this same combination
$\BdD - \BdMc/2$ and thereby restore the sensitivity.

\vspace{2mm}
\section*{Acknowledgments}

The authors would like to thank Junting Huang, Shun Zhou, Yue Meng, and Iwan Morton-Blake
for useful discussions.
The authors are supported by the National Natural Science
Foundation of China (12425506 and 12375101).
SFG is also an affiliate member of Kavli IPMU, University of Tokyo.

\addcontentsline{toc}{section}{References}
\bibliographystyle{utphysGe}
\bibliography{references}

\clearpage
\onecolumngrid
\begin{center}
    \textbf{\large Supplementary Material}
\end{center}
\setcounter{equation}{0}        
\renewcommand{\theequation}{S\arabic{equation}}

\suppsection{JUNO Analysis}
\label{sec:appA}

The reactor antineutrino signal is detected through
the inverse beta decay (IBD), 
$\bar\nu_e + p \to e^+ + n$, with the prompt positron energy
encoding the neutrino energy $E_\nu$. 
We analyze both the released JUNO data
\cite{JUNO:2025gmd, JUNO:2026talk} following the
real-data pipeline \cite{Esteban:2026phq}
while the projected sensitivities at the nominal exposure
are obtained following the Asimov approach \cite{Forero:2021lax}.

For both cases, the expected number of events in the 
$k$-th reconstructed-energy bin is evaluated as, 
\begin{equation}
N_k = C \sum_r \frac{\mathcal P_r}{4\pi L_r^2}\!\int\! dE_\nu\,
\phi(E_\nu)\,\sigma_{\rm IBD}(E_\nu)\,\langle P_{\bar e\bar e}\rangle_k\,
R_k(E_\nu),
\label{eq:Nk_events}
\end{equation}
The summation goes over the nine reactor cores
(six Yangjiang and two Taishan cores at
baselines $L_r\in[52.11,52.82]$\,km, plus the Daya~Bay complex at 
$L\simeq215$ km) with $\mathcal P_r$ being the core
thermal power weight, 
$\sigma_{\rm IBD}$ the IBD cross section \cite{Vogel:1999zy},
$R_k$ the detector energy-response function, and $C$
an overall normalization.
The effective single-baseline approximation gives $\langle L \rangle \approx 52.5$ km.

The flux $\phi(E_\nu)$ is the unoscillated 
$\bar\nu_e$ spectrum reconstructed by the Daya~Bay collaboration~\cite{DayaBay:2025ngb}.
We assume for all nine cores the burn-cycle-averaged composition reported by Daya~Bay,
\begin{equation}
f_{\Nuc{235}{U}} : f_{\Nuc{238}{U}} : f_{\Nuc{239}{Pu}} : f_{\Nuc{241}{Pu}}
= 0.564 : 0.076 : 0.304 : 0.056\,.
\label{eq:fission_fractions}
\end{equation}
We take the Huber--Mueller
flux \cite{Huber:2011wv,Mueller:2011nm},
\begin{equation}
\Phi(E_\nu) = \sum_{i} f_i\,\phi_i(E_\nu)\,,
\qquad i\in\{\Nuc{235}{U},\Nuc{238}{U},\Nuc{239}{Pu},\Nuc{241}{Pu}\}\,,
\label{eq:flux_isotopes}
\end{equation}
with the fission fractions of \geqn{eq:fission_fractions}.
The $\Nuc{235}{U},\Nuc{239}{Pu},\Nuc{241}{Pu}$ spectra follow
the $\beta$-conversion
fits of Ref.\,\cite{Huber:2011wv} and the non-fissile $\Nuc{238}{U}$ contribution
is taken \emph{ab initio} from Ref.\,\cite{Mueller:2011nm}.

Because the RG running enters through the detection momentum
transfer $Q^2$ which cannot be experimentally reconstructed,
the cross section is averaged over $Q^2$ for each energy bin
\cite{Ge:2024ibn},
\begin{equation}
\langle P_{\bar e\bar e}\rangle_k \equiv
\frac{\int_{Q^2_d\in k} dQ^2_d\,(d\sigma/dQ^2_d)\,P_{\bar e\bar e}(Q^2_d)}
 {\sigma_k}\,.
\label{eq:Pweighted_analysis}
\end{equation}
The two analyses differ in the treatment of the flux,
the energy response, and the backgrounds.

The prompt energy $E_{\rm pr} \equiv E_e+m_e$
is related to the neutrino energy $E_\nu$ by the IBD kinematics,
$E_{\nu}\approx E_{\rm pr} + 0.78$ MeV.
We use the energy-resolution reported by the JUNO collaboration
\cite{JUNO:2025fpc},
\begin{equation}
  \sigma(E_{\rm pr})
=
  E_{\rm pr}
  \sqrt{a^2/E_{\rm pr}+b^2},
\qquad
  a=3.3\%,\ \ b=1.0\%\,,
\label{eq:resol_data}
\end{equation}
with the positron non-linearity function $F_{\rm n.l.}$
extracted from the Fig.\,6d of Ref.\,\cite{JUNO:2025gmd}.
The spectrum is
divided into 66 bins over $E_\nu\in[1.8,8.0]$\,MeV, above which the
mass-ordering sensitivity saturates~\cite{Forero:2021lax}.

The matter effects are included through the effective
in-matter mixing parameters, expanding the dominant solar
term to linear order
in $A\equiv 2E_\nu V/\Delta m^2_{21}$ at a constant density
$2.55$\,g/cm$^3$~\cite{JUNO:2025gmd}.
The prior constraint on the reactor mixing angle
$\sin^2\theta_r = 0.022\pm0.00056$
\cite{DayaBay:2021dqj} is imposed as a Gaussian distribution.
The background splits into five components --- $\Nuc{9}{Li}/\Nuc{8}{He}$, geoneutrinos, world
reactors, $\Nuc{214}{Bi}/\Nuc{214}{Po}$, and others ---
read from the Fig.\,3 and
normalized to the Tab.\,1 of Ref.\,\cite{JUNO:2025gmd}.
The systematic uncertainties comprise
the signal normalization ($1.8\%$), the five background normalizations
($33\%, 42\%, 10\%, 56\%, 100\%$ for the aforementioned five
components, respectively), the $\Nuc{9}{Li}/\Nuc{8}{He}$
shape ($20\%$ at $1$\,MeV and
linear in energy), the $25$ flux uncertainties,
the two energy-scale uncertainties
$(\xi_{\rm scl},\xi_{\rm bias})$ at $0.5\%$,
and an energy-resolution uncertainty ($5\%$).

We adopt the Combined Neyman-Pearson (CNP) test statistic
\cite{Ji:2019yca} that is used by the collaboration,
\begin{equation}
  \chi^2_{\rm CNP}
=
  \sum_k \frac{(P_k-O_k)^2}{\sigma_k^2},
\qquad
  \sigma_k^2
=
  \frac{3}{1/O_k + 2/P_k},
\label{eq:chisq_CNP}
\end{equation}
with $O_k$ ($P_k$) being the observed (predicted) counts.
The Gaussian uncertainty penalties are added and profiled at
each point. The first release targets the slow mode and it carries
essentially no mass-ordering information on its own.

\vspace{1mm}
We adopt the nominal $2400$ effective detector-day exposure
($8$ years at $82\%$ live time, $26.6$\,GW$_{\rm th}$, $20$\,kton
fiducial mass) \cite{Forero:2021lax} to generate an Asimov
data set under the normal ordering (NO) with the NuFIT~6.1 best-fit parameters
and no running.
The fitting under the hypothesis of interest with the Pearson
statistic is given as,
\begin{equation}
  \chi^2(\vec X,\vec\xi)
=
  \sum_{k=1}^{N_{\rm bin}}
  \frac {[N_k^{\rm pred}(\vec X,\vec\xi)-N_k^{\rm data}]^2}
        {N_k^{\rm data}}
+ \sum_\alpha \frac{\xi_\alpha^2}{\sigma_\alpha^2}
+ \chi^2_{\rm NL},
\label{eq:chisq_proj}
\end{equation}
where $\vec X$ collects the RG running parameters
$\DTS,\DdMa,\DdD,\DdMc$. On the other hand, the nuisance parameters
$\vec\xi$ are the signal normalization,
the energy scale (a $0.7\%$ bias), the bin-to-bin flux-shape
uncertainty ($1\%$), and the five background normalizations rescaled to the
projected exposure; $\chi^2_{\rm NL}$ penalizes the non-linear energy response
\cite{Forero:2021lax}. The prediction depends on the uncertainties linearly,
\begin{eqnarray}
  N_k^{\rm pred}
=
  N_k^{\rm sig}(\vec X)
  \left( 1+\xi_{\rm norm}+\xi_{\rm Escale}f_k^E \right)
+ \sum_j B_j(1+\xi_j)s_j(k)+\xi_{\rm flux}\text{-terms},
\end{eqnarray}
with $f_k^E$ the discrete derivative of the signal spectrum under a uniform
energy shift. The mass-ordering sensitivity is quantified by
\begin{equation}
  \Delta\chi^2_{\rm IO}
\equiv
  \chi^2_{\min}[\text{IO}]
- \chi^2_{\min}[\text{NO}].
\label{eq:dchi2_IO_app}
\end{equation}
By construction with the Asimov data, $\chi^2_{\min}[\text{NO}]=0$,
such that the sensitivity is set by $\chi^2_{\min}[\text{IO}]$.
The IO fitting drives $|\Delta m^2_{ee}|^{\rm IO}$
and all the nuisance parameters to those values that best
mimics the NO wiggle pattern \cite{Forero:2021lax}. With
$\Delta m^2_{ee}=2.52\times10^{-3}$\,eV$^2$,
this gives $\Delta\chi^2_{\rm IO}=8.9$
in the standard case. The same machinery,
with the relevant parameter floated,
yields the projected sensitivity to the RG
running of each oscillation parameter.

\suppsection{RG Corrected Fast Oscillation Mode} 

We generalize the approach established in \cite{Minakata:2007tn} to
systematically include the RG running parameters.
For notation convenience,
we define the effective RG-corrected oscillation phases
in \protect\geqn{eq:Pee_RG_full} as
$\hat{\Phi}_{ij}\equiv \Phi_{ij}+\DP_{ij}$. 
These RG-corrected phases  inherently satisfy the cyclic relation 
$\hat{\Phi}_{32}+\hat{\Phi}_{21}=\hat{\Phi}_{31}$. 
Consequently, the standard effective atmospheric phase combination
$\Phi_{ee} \equiv \cS^2 \Phi_{31} + \sS^2\Phi_{32}$ extends to
\begin{align}
  \hat \Phi_{ee}
\equiv
  \cS^2 \hat{\Phi}_{31}
+ \sS^2\hat{\Phi}_{32}
=
  \cS^2 \Phi_{31}
+ \sS^2\Phi_{32}
+ (\cS^2\DP_{31} + \sS^2\DP_{32}).
\label{eq:ee_phase_RG_app}
\end{align}

Multiplying \geqn{eq:ee_phase_RG_app} by $4E_\nu/L$ yields the 
RG correction to the effective reactor atmospheric mass splitting: 
\begin{equation}
  \Delta \hat{m}^2_{ee}
\equiv
  \Delta m^2_{ee}
+ \dfrac{4E_\nu} L (\cS^2\DP_{31}+\sS^2\DP_{32})
\approx
  \Delta m^2_{ee}
- \dfrac{4E_\nu}{L} \DPee,
\end{equation}
where we have used \protect\geqn{eq:def_upsilon} in the second
equality. This reproduces the apparent shift of $\Delta m^2_{ee}$
quoted in the main text.
The effective phase combination
driven by the RG running is
defined as $\DPee \equiv \tfrac 1 2 \DdD + \tfrac 1 4 \cS^2 \DdMa - \tfrac 1 4 \DdMc$.
Thus, the effective mass splitting $\Delta m^2_{ee}$
inherits its apparent shift directly from the RG running of the fundamental oscillation phases.

Recall that the fast oscillation component of the probability 
in \protect\geqn{eq:Pee_RG_full} is given by 
\begin{align}
  P_{\bar{e}\bar{e}}^{\text{fast}}
\equiv
- \sin^2 (2 \TR)
\Big[
  \cS^2 (1+\DT_{31}) \sin^2 \hat{\Phi}_{31}
+ \sS^2 (1+\DT_{32}) \sin^2 \hat{\Phi}_{32}
\Big].
\end{align}
The fast oscillation mode explicitly depends on the two
effective phases $\hat{\Phi}_{31}$ and $\hat{\Phi}_{32}$. 
To simplify this dependency, 
we can reparameterize the system
by choosing $\hat{\Phi}_{ee}$ and $\hat{\Phi}_{21}$ as our new
basis. By virtue of \geqn{eq:ee_phase_RG_app} and the cyclic
relation, the original phases can be reexpressed as 
$\hat{\Phi}_{31}=\hat{\Phi}_{ee}+\sS^2 \hat{\Phi}_{21}$ 
and $\hat{\Phi}_{32}=\hat{\Phi}_{ee}-\cS^2 \hat{\Phi}_{21}$. 
Substituting these into the fast probability term yields:
\begin{equation}
  P_{\bar{e}\bar{e}}^{\text{fast}}
=
- \sin^2(2\TR)
\Big[
  \cS^2 (1+\DT_{31}) \sin^2 (\hat{\Phi}_{ee} + \sS^2 \hat{\Phi}_{21})
+ \sS^2 (1+\DT_{32}) \sin^2 (\hat{\Phi}_{ee} - \cS^2 \hat{\Phi}_{21})
\Big].
\end{equation}
By applying standard trigonometric identities and substituting
\geqn{eq:def_omega} alongside the approximation 
$\cS^2 \DT_{31}+\sS^2 \DT_{32} \approx 0$, we can expand
this expression into terms dependent on $2\hat{\Phi}_{ee}$:
\begin{align}
  P_{\bar{e}\bar{e}}^{\text{fast}}
= &
- \frac 1 2 \sin^2(2\TR)
\notag \\
&
+ \frac 1 2 \sin^2(2\TR)\cos(2\hat{\Phi}_{ee}) \left[\cS^2(1+\DT_{31}) \cos(2\sS^2\hat{\Phi}_{21})+\sS^2(1+\DT_{32}) \cos(2\cS^2\hat{\Phi}_{21}) \right]
\notag \\
&
- \frac{1}{2}\sin^2(2\TR) \sin(2\hat{\Phi}_{ee}) \left[\cS^2(1+\DT_{31}) \sin(2\sS^2\hat{\Phi}_{21})-\sS^2(1+\DT_{32}) \sin(2\cS^2\hat{\Phi}_{21}) \right].
\end{align}
The second and third terms are cosine and sine modulations of
$\hat{\Phi}_{ee}$, respectively. 
We can merge them into a single coherent cosine function, 
\begin{align}
P_{\bar{e}\bar{e}}^{\text{fast}}
= -\dfrac{1}{2}\sin^2(2\TR) \left[1- \sqrt{ 1 - \sin^2(2\TS)(1+\DT_{31}+\DT_{32}) \,\sin^2\hat{\Phi}_{21}} 
\,\cos(2\hat{\Phi}_{ee}+\hat{\Phi}_\odot)\right]+\mathcal{O}(\Delta^2) \,,  \label{eq:fast_term_expanded_app}
\end{align}
where we have introduced the RG-corrected solar phase
$\hat \Phi_\odot$,
\begin{equation}
  \tan\hat{\Phi}_{\odot}
\equiv
  \dfrac {\cS^2(1+\DT_{31}) \sin(2\sS^2\hat{\Phi}_{21})-\sS^2(1+\DT_{32}) \sin(2\cS^2\hat{\Phi}_{21}) }
         {\cS^2(1+\DT_{31}) \cos(2\sS^2\hat{\Phi}_{21})+\sS^2(1+\DT_{32}) \cos(2\cS^2\hat{\Phi}_{21})} \,.
\end{equation}

Extracting the explicit expression for $\hat{\Phi}_{\odot}$ 
directly from the trigonometric identities is algebraically cumbersome. 
A more efficient alternative is to map the terms to an auxiliary complex variable,
\begin{align}
  \hat Z
& \equiv
  \cS^2(1+\DT_{31}) e^{2i\sS^2 \hat{\Phi}_{21}} + \sS^2(1+\DT_{32}) e^{-2i\cS^2 \hat{\Phi}_{21}}
\notag \\
& =
  e^{-i \hat{\Phi}_{21} \cos 2\theta_{12}} \left[ \cS^2(1+\DT_{31}) e^{i \hat{\Phi}_{21}} + \sS^2(1+\DT_{32}) e^{-i \hat{\Phi}_{21}} \right],
\end{align}
The phase of $\hat{Z}$ directly corresponds to $\hat \Phi_\odot$,
\begin{equation}
  \hat{\Phi}_{\odot}
= \arg{\hat{Z}}
=
- \hat{\Phi}_{21}\cos2\TS
+ \arctan
\left(\dfrac{\cos2\TS+\cS^2\DT_{31}-\sS^2\DT_{32}}{1+\cS^2\DT_{31}+\sS^2\DT_{32}}\, \tan\hat{\Phi}_{21}\right) \,.
\end{equation}
Substituting Eqs.~(\ref{eq:def_omega}), and\,(\ref{eq:def_upsilon}), 
yields the more compact form,
\begin{equation}
  \hat{\Phi}_{\odot}
\approx
- \hat{\Phi}_{21}\cos2\TS+\arctan\left[(\cos 2\TS-\sin2\TS \,\DTS) \, \tan\hat{\Phi}_{21}\right] \,.
\end{equation} 

Since the RG running effects are perturbatively small,
we can perform a first-order Taylor expansion on the
arctangent function,
\begin{align}
  \hat{\Phi}_{\odot}
\approx &
- \Phi_{21}\cos 2\TS
+ \arctan(\cos 2\TS \tan\Phi_{21})
\notag \\
&
- \frac {\sin 2\TS \sin\Phi_{21}}
        {\cos^2\Phi_{21} + \cos^2 2\TS \sin^2\Phi_{21}}
\left[
  \cos\Phi_{21} \DTS
+ \frac 1 4 \cos 2\TS \sin 2\TS \sin\Phi_{21} \DdMa
\right].
\end{align}
The first two terms precisely reproduce the standard, uncorrelated solar phase 
$\Phi_\odot \equiv -\Phi_{21}\cos 2\TS + \arctan(\cos 2\TS \tan\Phi_{21})$ as derived in \cite{Forero:2021lax}. 
When incorporating the RG running, this solar phase receives
corrections from both the phase shift $\DdMa$
and the solar angle beta function $\DTS$. 
Notably, both of these parameters are directly probed 
by the slow oscillation sector. 

Finally, collecting the results from Eqs.\,(\ref{eq:ee_phase_RG_app})
and (\ref{eq:fast_term_expanded_app}) gives
the complete argument of the fast oscillation cosine:
\begin{align}
  \cos (2\hat{\Phi}_{ee}+\hat{\Phi}_\odot)
& \approx
\cos
\bigg\{
  2 \Phi_{ee}
+ \Phi_\odot
- (\DdD + \tfrac{1}{2}\cS^2\,  \DdMa-\tfrac{1}{2}
\DdMc) \notag \\
&
- \frac {\sin 2\TS \sin\Phi_{21}}
        {\cos^2\Phi_{21} + \cos^2 2\TS \sin^2\Phi_{21}}
  \left[
    \cos\Phi_{21} \DTS
  + \frac 1 4 \cos 2\TS \sin 2\TS \sin\Phi_{21} \DdMa
  \right]
\bigg\}.
\end{align}
Since $\DTS$ and $ \DdMa$ are tightly constrained by the slow sector limits, 
the dominant contribution to the overall phase shift 
stems from the RG-corrected effective phase combination 
$\hat{\Phi}_{ee}$, which is solely dependent on $\DPee\equiv\tfrac{1}{2}\DdD + \tfrac{1}{4}\cS^2\,  \DdMa-\tfrac{1}{4} \DdMc $. 

\end{document}